\documentclass[aps,prl,twocolumn,superscriptaddress,showpacs,floatfix,nobibnotes]{revtex4}
\usepackage{epsfig}
\usepackage{epstopdf}
\usepackage{CJKfntef}

\usepackage{graphicx}
\usepackage{longtable}
\usepackage{CJK}
\usepackage{color}

\usepackage{mathptmx, courier, pifont}
\usepackage[scaled=0.92]{helvet}
\usepackage[T1]{fontenc}
\usepackage{textcomp}

\begin{document}

\begin{CJK*}{GBK}{song}

\title{A Note on the Consistent-$Q$ Scheme for Odd-Odd Nuclei}

\author{Xiao-Tong Li}
\affiliation{Department of Physics, Liaoning Normal University,
Dalian 116029, P. R. China}

\author{Xi Deng}
\affiliation{Department of Physics, Liaoning Normal University,
Dalian 116029, P. R. China}

\author{Yu Zhang }\email{dlzhangyu_physics@163.com}
\affiliation{Department of Physics, Liaoning Normal University,
Dalian 116029, P. R. China}

\date{\today}

\begin{abstract}
Compared with even-even nuclear systems, the dynamical structure of low-lying excited states in odd-odd nuclei poses significantly greater theoretical challenges. The interacting boson fermion fermion model provides an effective framework for capturing the structural properties of odd-odd nuclei. Within this algebraic framework, the consistent-$Q$ scheme, which was widely used to map nuclear shape phase diagram, is extended to odd-odd systems to examine how unpaired nucleons influence level structures and their evolution across prototypical transitional regions. The results demonstrate that the presence of unpaired nucleons does not suppress the critical behavior characterizing level evolution across the U(5)-SU(3), U(5)-O(6) and SU(3)-O(6) shape transitions, suggesting that shape phase transitions remain a fundamental mechanism governing the low-lying structural evolution of odd-odd nuclei in the heavy and intermediately-heavy mass regions.
\end{abstract}


\maketitle

\end{CJK*}

\begin{center}
\vskip.2cm\textbf{1. Introduction}
\end{center}\vskip.2cm

Shape phase transitions (SPTs) are an important way for the structural evolution of nuclei in the heavy and intermediately-heavy mass regions, and related topics have been extensively investigated over the past two decades~\cite{Casten2007,Cejnar2009,Cejnar2010}. However, in contrast to the wealth of theoretical and experimental studies on SPTs in even-even and odd-A nuclei~\cite{Casten2007,Cejnar2009,Cejnar2010,Jolie2004,Iachello2005,Alonso2005,Alonso2006,LML2007,Alonso2007,Alonso2007II,Alonso2009,Boyukata2010,
Iachello2011,Petrellis2011,Zhang2013,Zhang2013II,Zhang2015,Nomura2016,Bucurescu2017,Nomura2017,Nomura2017II,Zhang2018,
Quan2017,Nomura2018,Bucurescu2018,Nomura2020}, there are few theoretical analyses for odd-odd nuclei. Existing work on odd-odd nuclei is mostly confined to phenomenological descriptions of ground-state properties for specific cases~\cite{Zhang2013,Bucurescu2017,Zhang2021}. One possible reason lies in the intricate interplay between single-particle and collective excitations, which makes the low-lying excitation spectra of odd-odd nuclei significantly more complex. In particular, pronounced shape fluctuations near the critical point of an SPT further complicate theoretical modeling. The interacting boson model (IBM)~\cite{IachelloBook87} is one of the most convenient and effective frameworks for analyzing SPTs in even-even nuclei~\cite{Cejnar2009}. In the IBM, typical SPTs can be characterized by a two-parameter consistent-$Q$ Hamiltonian~\cite{Warner1983}, and the nuclear shape-phase diagram can be accordingly mapped onto a triangle (referred to as Casten triangle). Extended to odd-A and odd-odd systems, the IBM
generalizes to the interacting boson fermion model (IBFM) and the interacting boson fermion fermion model (IBFFM), respectively~\cite{IachelloBook91}.
Within the framework of IBFM, an odd-A nucleus can be approximately described as a coupled system comprising an even-even bosonic core plus an unpaired fermion (either particle
or hole)~\cite{IachelloBook91}. In such systems, SPTs can be explored as quantum phase transitions (QPTs) between two distinct dynamical symmetry limits of the bosonic core~\cite{Iachello2011}, provided that the
ground-state shape is predominantly governed by the core deformation. Similarly, within the framework of the IBFFM, an odd-odd nucleus can be treated as an even-even core coupled to two unpaired fermions.

Previous IBFM-based analysis of SPTs~\cite{Zhang2018,Zhang2021} have demonstrated that the unpaired (odd) particle can influence various types of SPTs in qualitatively different ways~\cite{Boyukata2010,Iachello2011,Petrellis2011}. These influences manifest clearly in the evolution of low-lying level energies and characteristic energy ratios~\cite{Petrellis2011}, which are sensitive to ground state deformations. In this work, we extend such analyses to odd-odd nuclei within the IBFFM framework. Specifically, our investigation focuses on dynamical correlations between the IBM and IBFFM models, and their impact on yrast level systematics.

\begin{center}
\vskip.2cm\textbf{2. The Model}
\end{center}\vskip.2cm

The Hamiltonian and physical operators in the IBM framework are constructed from two kinds of
boson operators: a scalar $s$-boson with $J^\pi=0^+$ and a quadrupole
$d$-boson with $J^\pi=2^+$~\cite{IachelloBook87}. The IBM contains three dynamical symmetry (DS) limits, each characterized by a distinct subgroup chain of the U(6)
group:
\begin{eqnarray}\nonumber
&&\mathrm{U(6)} \supset \mathrm{U(5)} \supset \mathrm{SO(5)} \supset \mathrm{SO(3)}\, ,\\
&&\mathrm{U(6)} \supset \mathrm{O(6)} \supset \mathrm{SO(5)} \supset
\mathrm{SO(3)}\, ,\\ \nonumber&&\label{SU3} \mathrm{U(6)} \supset
\mathrm{SU(3)} \supset \mathrm{SO(3)}\, .
\end{eqnarray}
The widely used consistent-$Q$ Hamiltonian~\cite{Warner1983} can be parameterized as
\begin{equation} \label{IBM}
\hat{H}_\mathrm{B}=\varepsilon \left[ (1-\eta)\hat{n}_{d} -
\frac{\eta}{4N}\hat{Q}_{\mathrm{B}}\cdot \hat{Q}_{\mathrm{B}}
\right]\, ,
\end{equation}
where $N$ is the total boson number, and $\hat{n}_d= \sqrt{5}(d^\dag \times\tilde{d})^{(0)}$
with $\tilde{d}_u=(-1)^ud_{-u}$ is the $d$ boson number, representing a pairing-like interaction. The quadrupole operator is given by
\begin{equation}
\hat{Q}_\mathrm{B}^{\chi}=(d^{\dag} s + s^{\dag} \tilde{d})^{(2)} +
\chi (d^{\dag}\times\tilde{d})^{(2)}\, .
\end{equation}
Finally, $\eta$ and $\chi$ are two control parameters, satisfying $\eta\in[0,~1]$ and $\chi\in[-\sqrt{7}/2,~0]$, and $\varepsilon$
is an overall scale factor, which is set to be unity in all subsequent discussions for convenience.

\begin{figure}
\begin{center}
\includegraphics[scale=0.35]{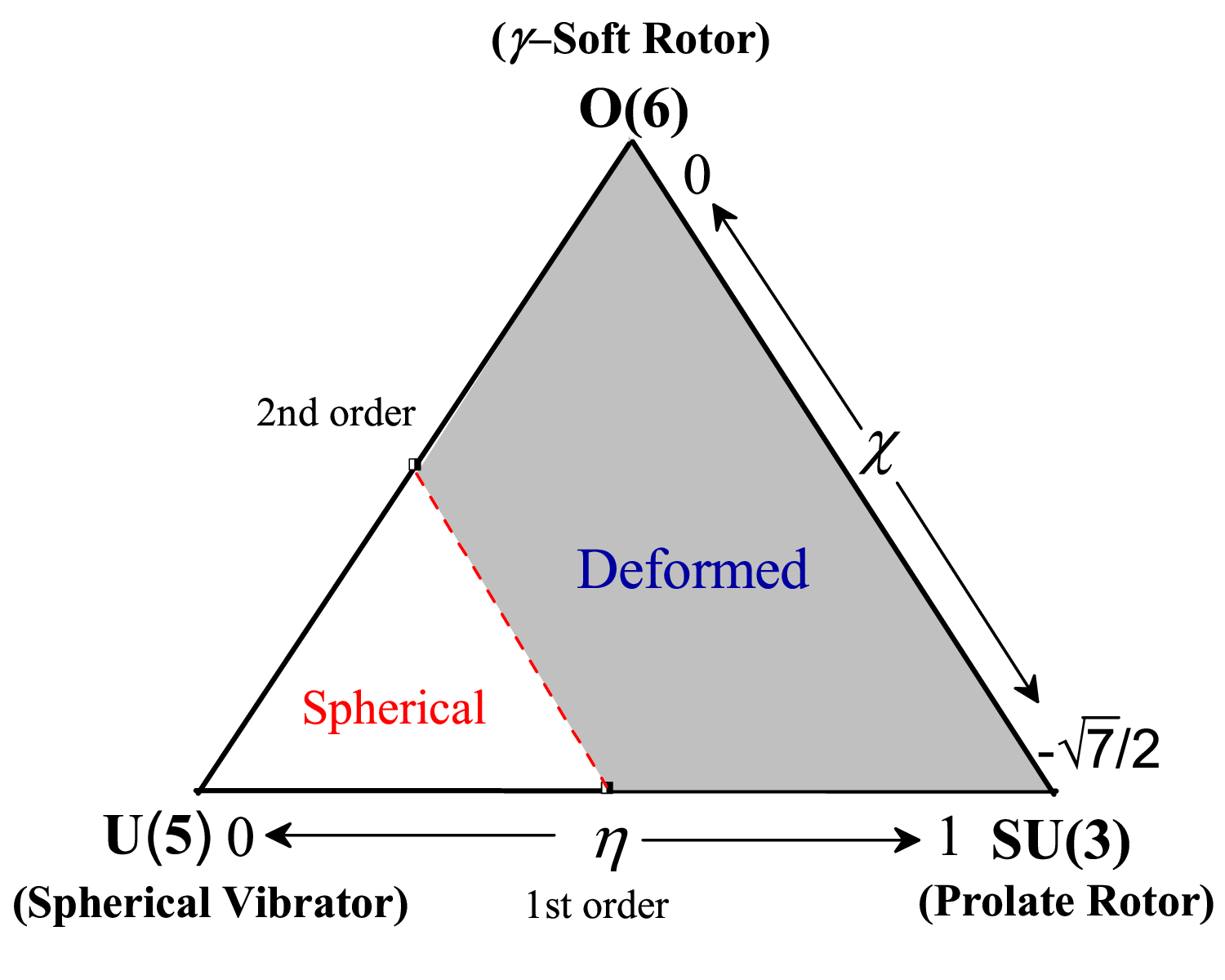}
\caption{(Color online) The schematic shape phase diagram described by the consistent-$Q$ Hamiltonian in (\ref{IBM}).  \label{F1}}
\end{center}
\end{figure}

With this consistent-$Q$ Hamiltonian, the shape phase diagram in the $\eta\times\chi$ parameter space is shown in Fig.~\ref{F1}. One can find that the three symmetry limits of the IBM locate at the vertices of the triangle: the U(5) limit at $(\eta,\chi)=(0,0)$, the O(6) limit at $(\eta,~\chi)=(1,~0)$, and the SU(3) limit at $(\eta,~\chi)=(1,~-\sqrt{7}/2)$. The corresponding quadrupole geometry (shape) and associated collective mode in each limit can be deduced from the mean-field potential expressed in terms of the quadrupole deformation parameters $(\beta,\gamma)$, using the coherent-state method~\cite{IachelloBook87}. Specifically, the mean-field calculations yield: (i) a $\gamma$-independent potential with $\beta_\mathrm{min}=0$ for U(5), indicating spherical vibrator; (ii) a $\gamma$-independent potential with $\beta_\mathrm{min}>0$ for O(6), corresponding to $\gamma$-soft rotor; (iii) an axially symmetric potential with $\beta_\mathrm{min}>0$ and $\gamma_\mathrm{min}=0$ for SU(3), signifying a prolate rotor. Here, $(\beta_\mathrm{min},\gamma_\mathrm{min})$ denote the location of the global minimum of the mean-field potential. At the mean-field level, it has been rigorously established that the first-order SPTs between spherical and deformed shapes occur within the triangle phase diagram~\cite{Iachello2004}. In the large-$N$ limit, the critical line is analytically described by
$\eta_\mathrm{c}=\frac{14}{28+\chi^2}$ with $\chi\in[-\sqrt{7}/2,0)$. As clearly illustrated from Fig.~\ref{F1}, this critical line partitions the triangle into two distinct regions: the spherical and the deformed. Furthermore, a second-order transition takes place along the U(5)-O(6) leg ($\chi=0$) at the critical point $\eta_\mathrm{c}=0.5$, marking the spherical to $\gamma$-soft shape evolution. In contrast, the SU(3)-O(6) leg ($\eta=1$) is found to represent a crossover in the large-$N$ limit. Collectively, the triangle phase diagram provides a unified description of all characteristic quadrupole shapes and their associated shape-transitional regimes..

The consistent-$Q$ Hamiltonian given in (\ref{IBM}) can be solved by diagonalizing it within
the SU(3) basis $|N (\lambda,\mu)\tilde{\chi} LM_L\rangle$~\cite{Rosensteel1990}, which is characterized by the group chain~\cite{IachelloBook87}
\begin{equation}
\left |\begin{array}{cc}
\mathrm{U(6)}\supset \mathrm{SU}(3)\supset
\mathrm{SO}(3)\supset\mathrm{SO}(2)\\
N~~~~~~(\lambda,\mu)\tilde{\chi}~~~~~~~~L~~~~~~~~~~~~M_L\end{array}
\right \rangle\, ,
\end{equation}
where $(\lambda,\mu)$ lalbels the SU(3) irreducible representation, and $\tilde{\chi}$ denotes the
multiplicity index of the angular momentum $L$ (with $M_L$ being its magnetic quantum number) within that SU(3) representation $(\lambda,\mu)$.
For odd-odd systems, the consistent-$Q$ Hamiltonian takes the form
\begin{equation} \label{IBFFM}
\hat{H}_{\mathrm{BF}}=\varepsilon \left[ (1-\eta)\hat{n}_{d} -
\frac{\eta}{4N}\hat{Q}_{\mathrm{BF}}\cdot \hat{Q}_{\mathrm{BF}}
\right] \, ,
\end{equation}
with \begin{equation}\label{Q}
\hat{Q}_{\mathrm{BF}} =
\hat{Q}_\mathrm{B}^\chi+\hat{q}_\pi+\hat{q}_\nu\,
\end{equation}
and
\begin{eqnarray}
\hat{q}_\pi&=&\langle j_\pi\parallel Y^{(2)}\parallel j_\pi\rangle(\hat{a}_{j_\pi}^\dag\times\tilde{a}_{j_\pi})^{(2)},\\
\hat{q}_\nu&=&\langle j_\nu\parallel Y^{(2)}\parallel j_\nu\rangle(\hat{a}_{j_\nu}^\dag\times\tilde{a}_{j_\nu})^{(2)}\, ,
\end{eqnarray}
where $j_\pi$ and $j_\nu$ represent the angular momentum quantum number of the unpaired proton and neutron, respectively.
For simplicity, we have assumed here that the unpaired particle is confined to a single-$j$ orbit.
To obtain the eigenvalues and
eigenfunctions, we diagonalize the IBFFM Hamiltonian in the weak-coupling SU(3) basis~\cite{Zhang2021},
\begin{eqnarray}\label{su3}
&&|\alpha J M_J\rangle\equiv|N (\lambda,\mu)\tilde{\chi} L; (j_\pi j_\nu)j_{\pi\nu}; JM_J\rangle\\ \nonumber
&&=\sum_{L,M_L,j_{\pi\nu},m_{j_{\pi\nu}}}^{m_{j_{\pi}},m_{j_{\nu}}}\langle j_\pi m_{j_\pi},j_\nu m_{j_\nu}|j_{\pi\nu}m_{j_{\pi\nu}}\rangle\\ \nonumber
&&\times\langle LM_L,j_{\pi\nu}m_{j_{\pi\nu}}|JM_J\rangle|N(\lambda,\mu)\tilde{\chi}LM_L\rangle|j_\pi m_{j_\pi}\rangle|j_\nu m_{j_\nu}\rangle\, ,
\end{eqnarray}
where $L,~j_{\pi(\nu)},~j_{\pi\nu},~J$ represent the angular momentum for the bosonic core, the odd
proton (neutron), the coupled proton-neutron pair, and the total system, respectively, with the corresponding magnetic
quantum numbers denoted in order by $M_L$, $m_{j_{\pi(\nu)}}$, $m_{j_{\pi\nu}}$ and $M_J$.

\begin{center}
\vskip.2cm\textbf{2. Numerical Results and Discussions}
\end{center}\vskip.2cm

\begin{figure}
\begin{center}
\includegraphics[scale=0.35]{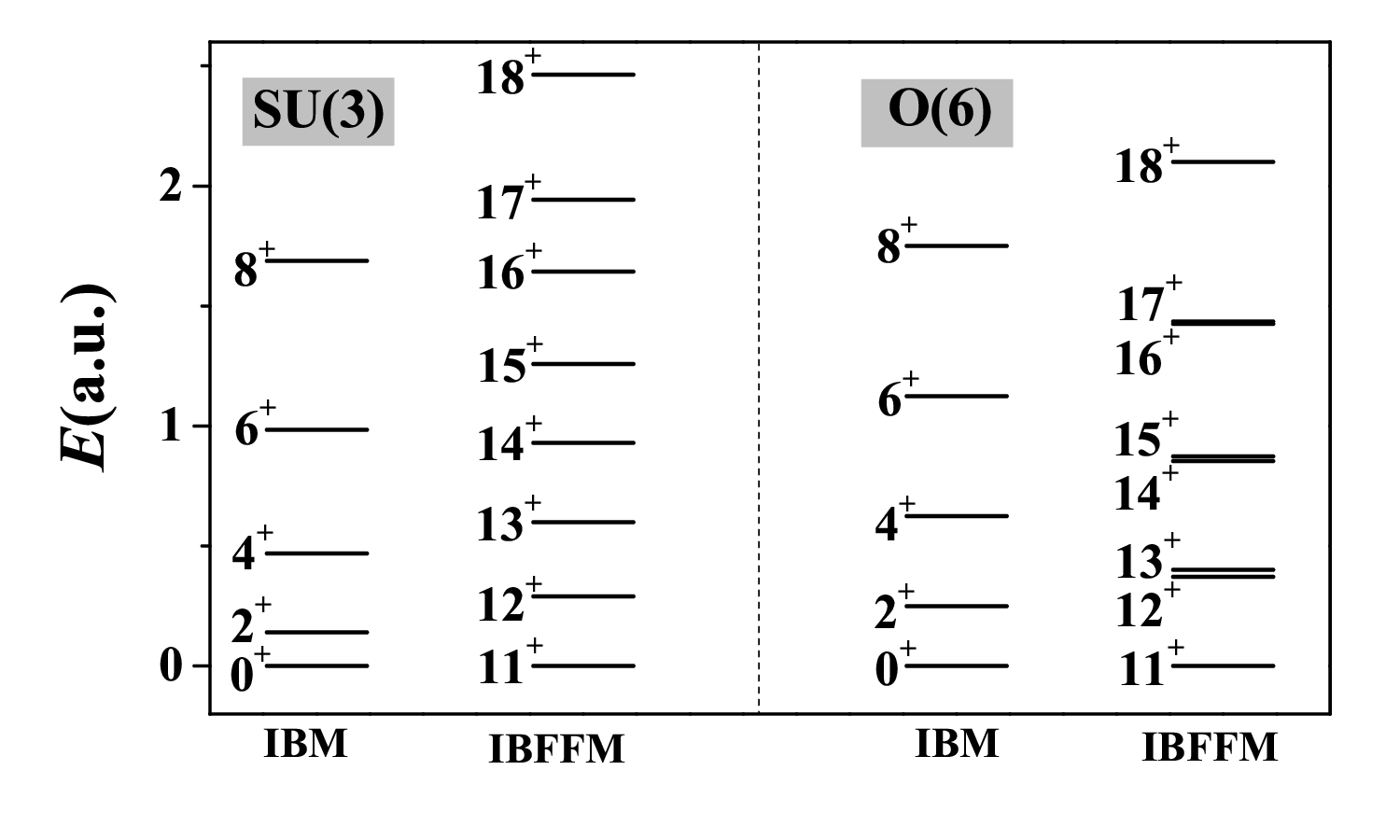}
\caption{(Color online) The yrast-state spectra (in any units) in the SU(3) limit and the O(6) limit. The results are solved from the consistent-$Q$
Hamiltonian given in (\ref{IBM}) and (\ref{IBFFM}), respectively. \label{F2}}
\end{center}
\end{figure}

To make a comparison between IBM and IBFFM, we compute the level energies using the corresponding consistent-$Q$ Hamiltonian with $N=4$.
First, we compare the yrast-state energy level structures of the IBM and IBFFM under the SU(3) limit $(\eta=1,\chi=-\sqrt{7}/2)$ and the O(6) limit $(\eta=1,\chi=0)$.
In the IBFFM, it is assumed that both odd proton and odd neutron occupy a $j=11/2$ single-particle orbit. Note that fixing the single-particle configuration is a common assumption in the theoretical analysis of high-spin spectra in odd-odd nuclei. As seen from Fig.~\ref{F2}, in the SU(3) limit, the IBM yrast state yields an axially-symmetric rotor spectrum, whose excitation energies follow the $J(J+1)$ rule. When two odd nucleons are coupled, the IBFFM also reproduced a well-defined rotational band structure in the SU(3) limit, where the total angular momentum of the ground state is given by $J_g=11$, resulting from the parallel alignment of the angular momenta of two unpaired nucleons. In contrast, in the O(6) limit, the yrast levels exhibit a triaxial rotational band structure with $E(4_1^+)/E(2_1^+)=2.5$. This triaxiality is particularly pronounced in the IBFFM, where adjacent levels form nearly degenerate pairs, clustering as $J^\pi=\{11^+\},~\{12^+,~13^+\},~\{14^+, 15^+\},\cdots$.
Despite with this distinction, the ground-state spin also arise from the parallel alignment of the angular momenta of two unpaired nucleons. It should be noted that, in the U(5) limit, the consistent-$Q$ formulism yields identical level patterns in between the IBM and IBFFM, except that levels in the latter exhibit a high degree of degeneracy. This behavior is readily understood by examining the IBFFM Hamiltonian given in Eq.~(\ref{IBFFM}), which, like its IBM counter part (see Eq.~(\ref{IBM})), contains only the $\hat{n}_d$ term in the U(5) limit.

\begin{figure}
\begin{center}
\includegraphics[scale=0.25]{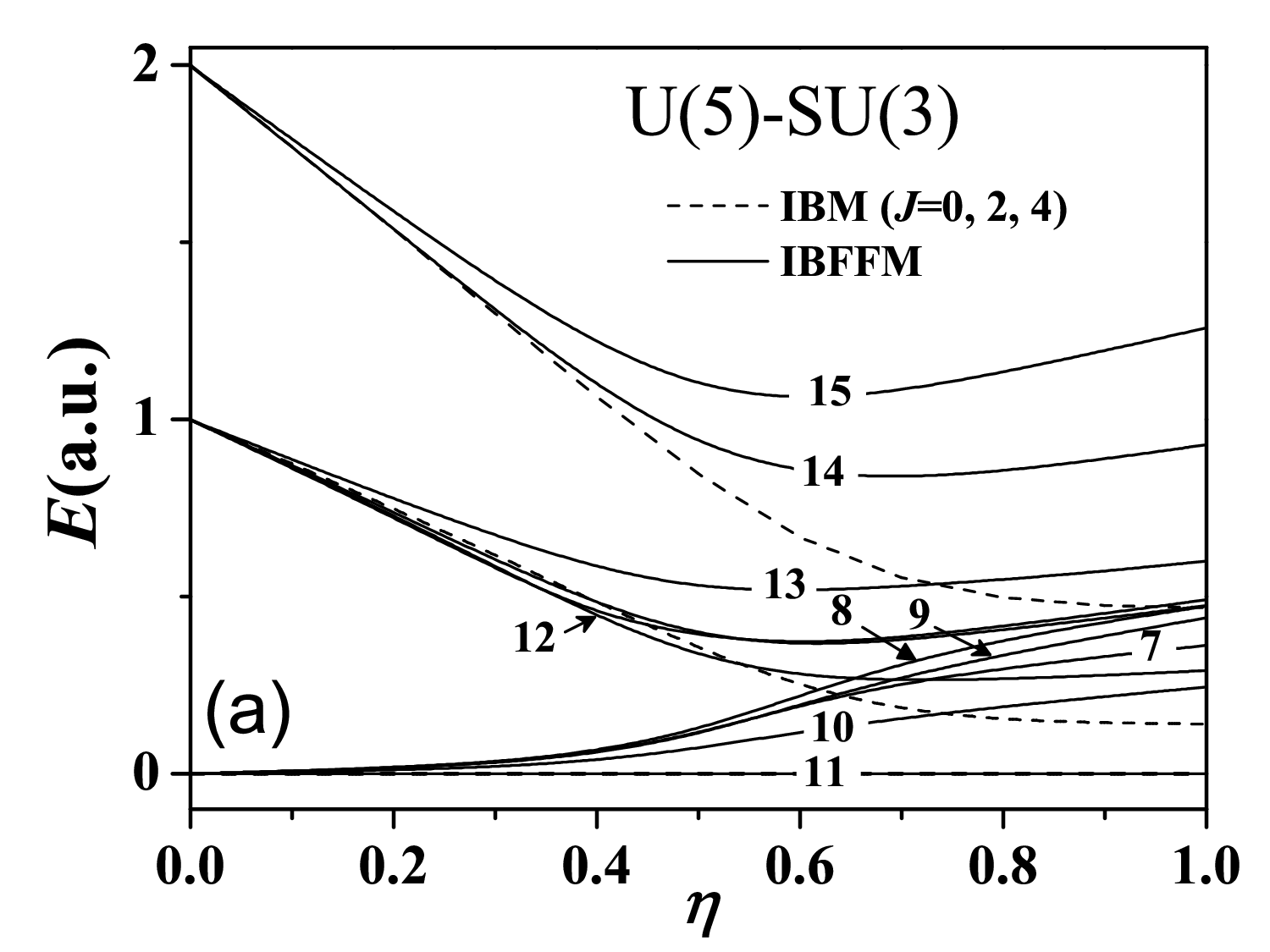}
\includegraphics[scale=0.25]{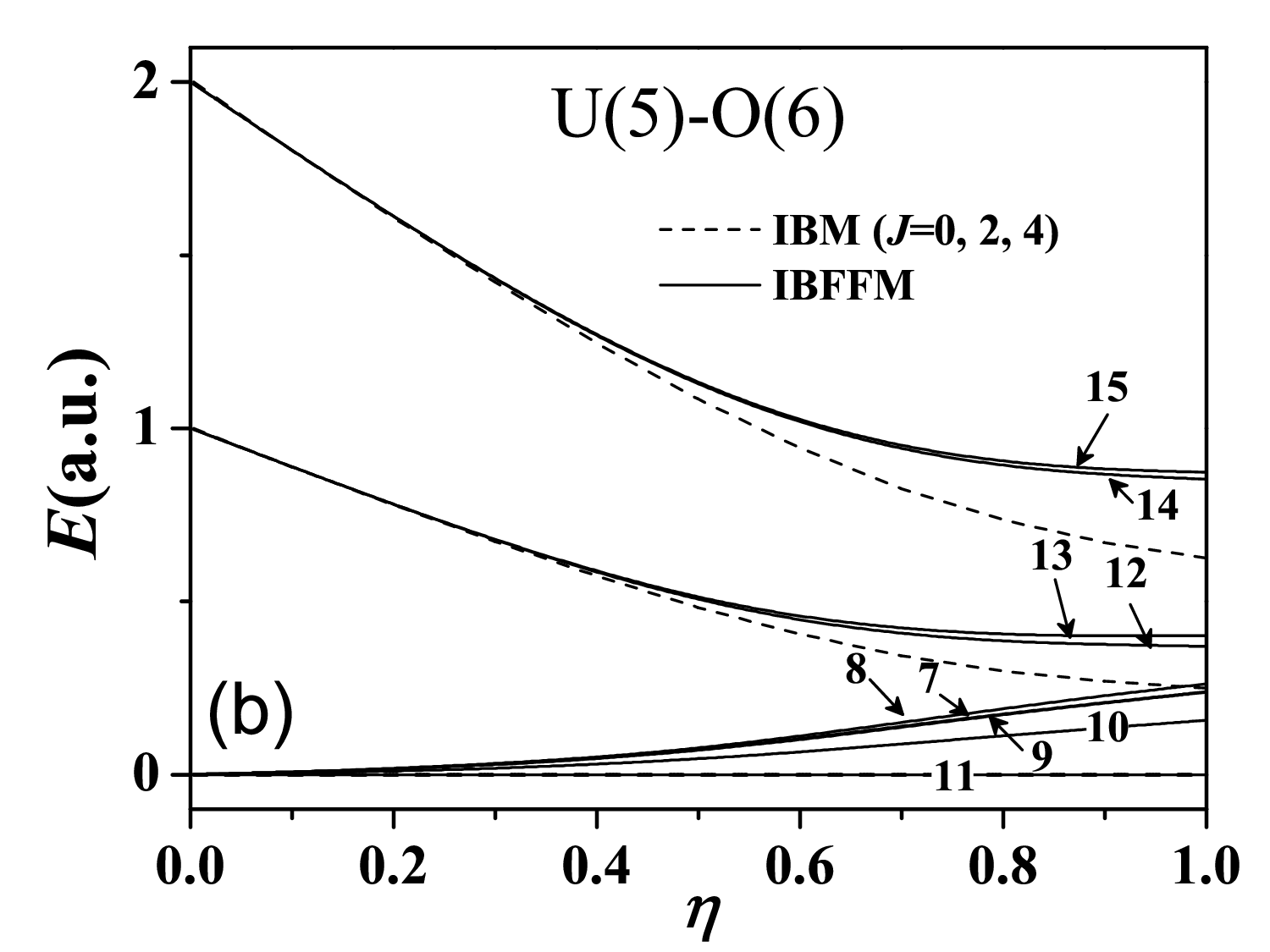}
\includegraphics[scale=0.25]{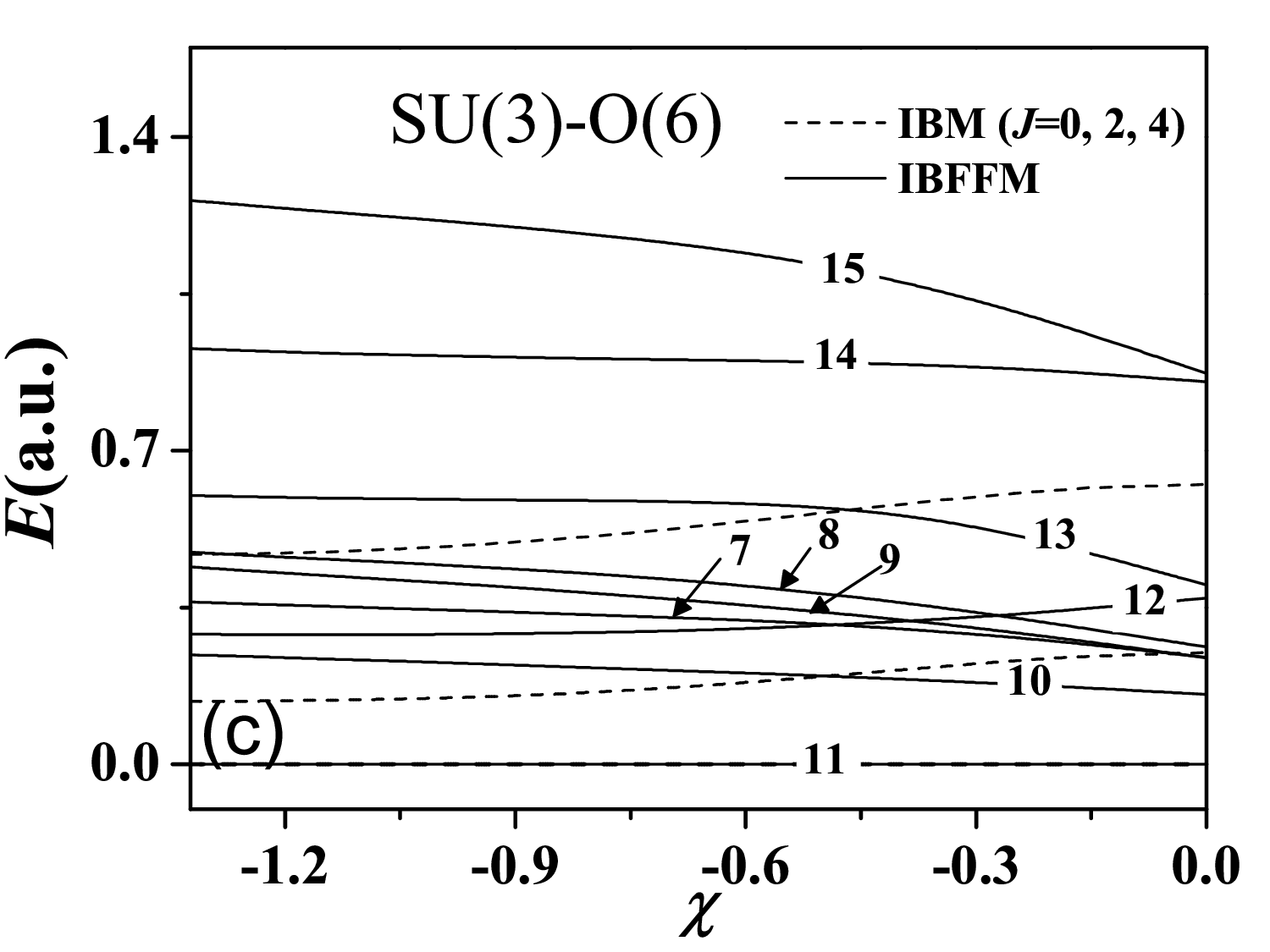}
\caption{(Color online) (a) The yrast level evolutions in the U(5)-SU(3) transitional region with the dashed lines denoting the IBM results for $J=0,~2,~4$
and the solid lines representing the IBFFM results for $J=7-15$. (b) Same as in (a) but for the U(5)-O(6) transition. (c) Same as in (a) but for the SU(3)-O(6) transition. \label{F3}}
\end{center}
\end{figure}

To analyze the transitional behaviors among the U(5), SU(3) and O(6) limits, we examine the evolution of yrast-state energy levels across the corresponding transitional region. As shown in Fig.~\ref{F3}(a), the IBFFM yrast energy levels in the U(5) limit form degenerate multiples, as expected. Specifically, the yrast states with $J=8-11$ are degenerate and share the same energy as the $0_1^+$ state in the IBM, while those with $J=12-13$ and $J=14-15$ are energetically identical to the $2_1^+$ and $4_1^+$ states in the IBM, respectively. Since the Hamiltonian in the U(5) limit contains no single-particle contributions, these degeneracies arise naturally from angular-momentum coupling considerations. For example, in the U(5) limit, one-phonon excited yrast states formed by coupling the two-particle angular momentum to the core angular momentum uniquely yield total angular momenta $J=12-13$. This is because the zero-phonon state cannot generate such $J$ values, while although two-phonon excitations can produce $J=12-13$, they do not correspond to yrast states. Furthermore, as evident from Fig.~\ref{F3}(a), the IBFFM yrast levels during the U(5)-SU(3) SPT evolve from the degenerate pattern characteristic of the U(5) limit toward a broken-degeneracy structure that reorganizes into rotational bands in the SU(3) limit. Particularly, certain high angular momentum yrast states (e.g., $J=13-15$) exhibit non-monotonic behavior near the critical point $\eta_\mathrm{c}=0.5$, which can be regarded as finite-$N$ precursors of the U(5)-SU(3) SPT. In contrast, the results shown in Fig.~\ref{F3}(b) and Fig.~\ref{F3}(c) indicate that evolution of level energies across the U(5)-O(6) and SU(3)-O(6) transitional regions is predominantly monotonic. Nevertheless, the characteristic evolutional features of yrast levels in the IBM are robustly preserved in the IBFFM. In other words, the transitional behavior remains essentially unaltered upon coupling two unpaired fermions to the bosonic core.

\begin{figure}
\begin{center}
\includegraphics[scale=0.25]{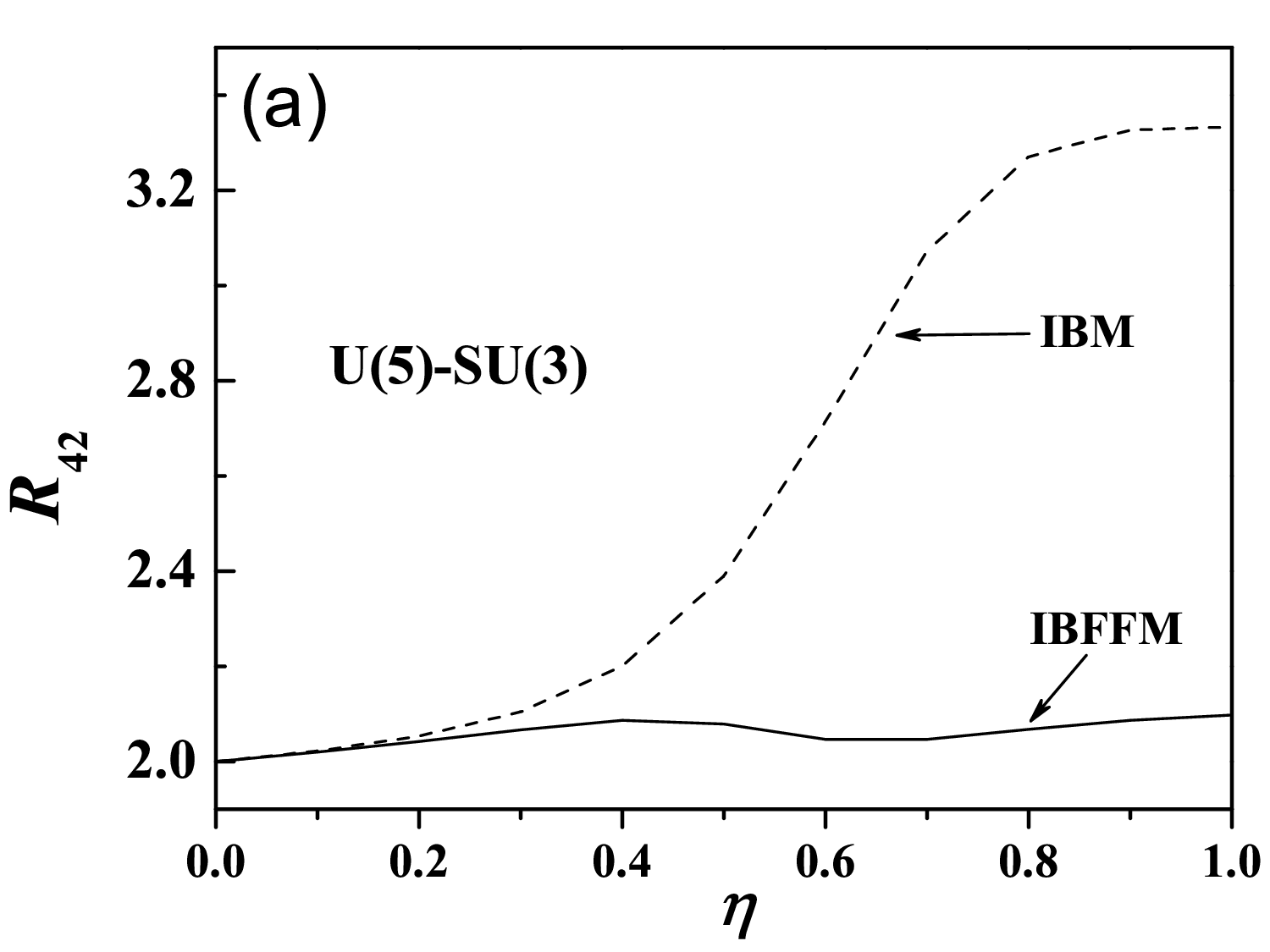}
\includegraphics[scale=0.25]{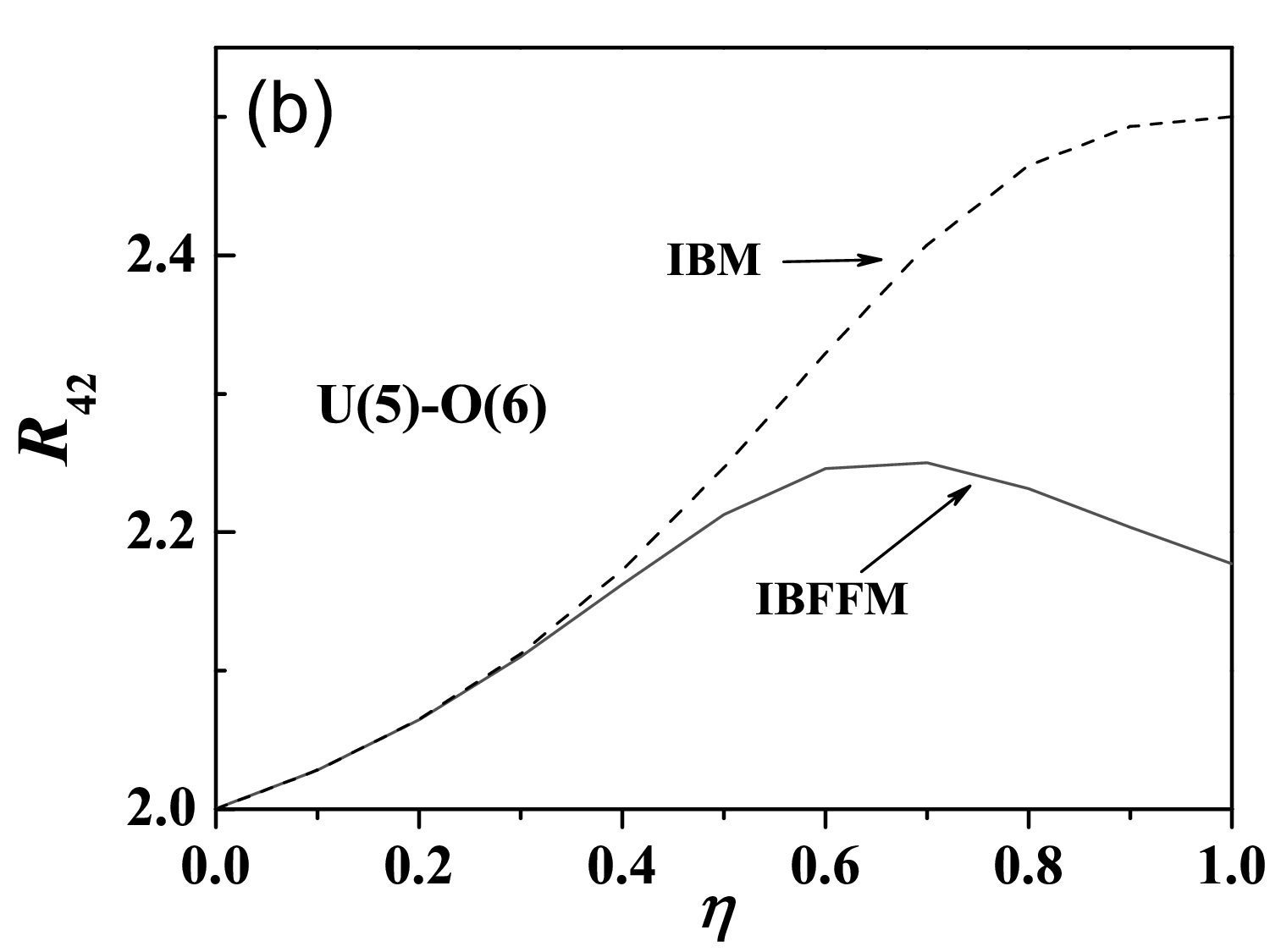}
\includegraphics[scale=0.25]{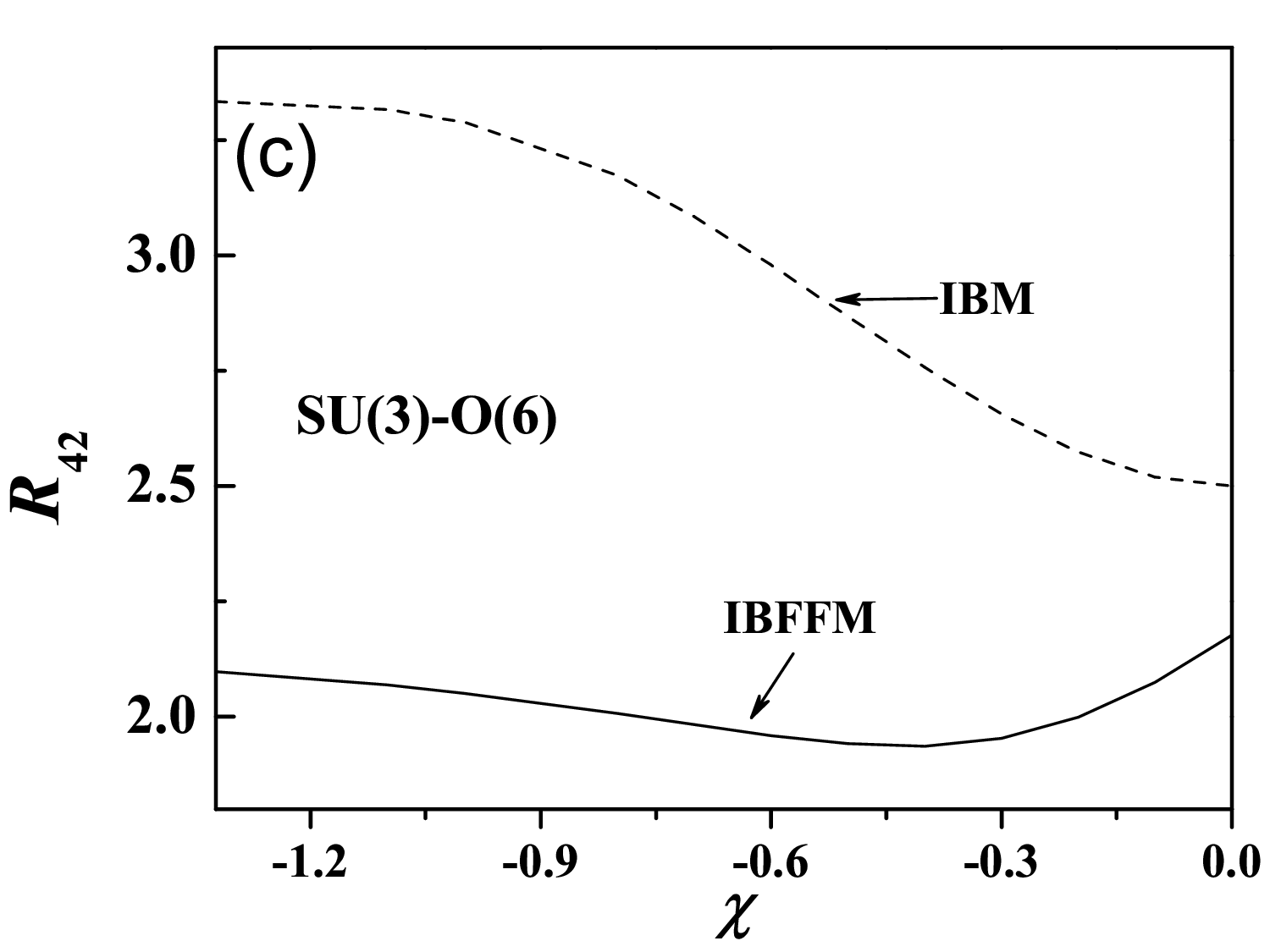}
\caption{(Color online) (a) The evolutions of the energy ratios $R_{4/2}$ (defined in the text) in the U(5)-SU(3) transitional region for the IBM and the IBFFM. (b) Same as in (a) but for the U(5)-O(6) transition. (c) Same as in (a) but for the SU(3)-O(6) transition. \label{F4}}
\end{center}
\end{figure}

Existing analyses~\cite{Zhang2021} have demonstrated that typical energy ratios can serve as effective order parameters for identifying distinct transitions in even-even nuclei and odd-A nuclei. Among them, a simple yet effective order parameter is the energy ratio $E(4_1^+)/E(2_1^+)$. To facilitate a comparison with the results between the IBM and IBFFM, we define the energy ratio as $R_{4/2}=E((J_g+4)_1^+)/E((J_g+2)_1^+)$, with $J_g=0$ for the IBM and $J_g=11$ for the IBFFM. Fig.~\ref{F4} presents the evolution of $R_{4/2}$ across various transitional region. As shown in Fig.~\ref{F4}(a), the behavior of $R_{4/2}$ clearly signals the U(5)-SU(3) transition in the IBM system. Specifically, the ratio values rise sharply from $R_{4/2}=2.0$ to $R_{4/2}=3.3$ near the critical point $\eta=0.5$, reflecting that the system undergoes a phase transition from spherical vibrational mode to axially deformed rotational mode. In contrast, the corresponding evolution of $R_{4/2}$ in the IBFFM exhibits significant compression of this transition signature relative to its IBM counterpart. It means that this yrast spectral ratio is not a robust indicator for the U(5)-SU(3) transition in odd-odd nuclei. Fig.~\ref{F4}(b) reveals that the U(5)-O(6) phase transition can be identified in both the even-even and odd-odd systems via the evolution of $R_{4/2}$. However, the transitional amplitude is evidently reduced in the IBFFM. Similarly, the results in Fig.~\ref{F4}(c) suggest that $R_{4/2}$ is also less sensitive to the SU(3)-O(6) transition in the IBFFM compared to the IBM, where the ratio drops dramatically from $R_{4/2}=3.33$ to $R_{4/2}=2.5$ across this transitional region. All these results indicate that shape phase transitions, though present in the complex spectra of odd-odd nuclei, are difficult to discern from the evolution of simple energy ratios as commonly employed for even-even nuclei, owing to the influence of two unpaired fermions. Therefore, it is essential to identify alternative effective order parameters capable of revealing SPT signatures in odd-odd nuclei.

\begin{center}
\vskip.2cm\textbf{3. Summary}
\end{center}\vskip.2cm

SPTs constitute a fundamental mechanism driving the evolution of nuclear structure. While extensive theoretical investigations have been conducted on even-even and odd-A nuclei, systematic theoretical analyses of SPTs in odd-odd nuclei remain lacking. However, clear SPT signals can be unambiguously identified through the evolution of ground state properties in odd-odd nuclei, such as the odd-even binding energy difference~\cite{Zhang2013}. Within the algebraic framework, we conduct an analysis of the U(5)-SU(3), U(5)-O(6) and SU(3)-O(6) phase transitions, assuming two unpaired nucleons occupying fixed single-$j$ orbits. Our principal finding is that all the shape transitions remain robust in finite odd-odd nuclei, and the key features of level evolution identified in the even-even systems are well preserved in the odd-odd systems following the coupling of two unpaired odd nucleons. In contrast to the earlier approach~\cite{Zhang2013}, which relied solely on ground-state property variations to identify shape-phase transitions in odd-odd nuclei, the current results demonstrate that such transitions can also be reliably probed via the evolution of high-spin excitation spectra, under fixed single-particle configurations. Collectively, this study extends the investigation of nuclear shape-phase transitions to more complex odd-odd systems and offers a fresh perspective on the evolution of nuclear structure.

The present analysis employs the constant-$Q$ form of the Hamiltonian. Although this parametrization has been well validated for describing shape-phase transitions in even-even nuclei, its quantitative application to odd-odd nuclei may require inclusion of additional terms beyond the quadrupole-quadrupole interaction, such as the boson-fermion exchange term or the monopole term.
Related work is in progress.

\bigskip

\begin{acknowledgments}
Support from the National Natural Science Foundation of China
(12375113) is acknowledged.
\end{acknowledgments}



\end{document}